\input phyzzx.tex
\tolerance=1000
\voffset=-0.3cm
\hoffset=1.0cm
\def\rl{\rightline}

\def\t1{{\tilde 1}}

\def\t m^2{\tilde m^2}

\REF{\MAL} {J. M. Maldacena, hep-th/9711200.}
\REF{\KLEB} {S. S. Gubser, I. R. Klebanov and A. W. Peet hep-th/9602135.}
\REF{\kleb2} {I. R. Klebanov, hep-th/9702076}
\REF{\kleb3} {S. S. Gubser, I. R. Klebanov and A. A. Tseytlin,
hep-th/9703040.}
\REF{\kleb4} {S. S. Gubser and I. R. Klebanov,  hep-th/9708005.}
\REF{\MS} {J. M. Maldacena and A. Strominger, hep-th/9710014.}
\REF{\POL} {A. M. Polyakov,  hep-th/9711002.}
\REF{\GKP} {S.S. Gubser, I.R. Klebanov and A.M. Polyakov,
hep-th/9802109.}
\REF{\WIT} {E. Witten, hep-th/9802150.}
\REF{\0}  {K. Sfetsos and K. Skenderis, hep-th/9711138.}
\REF{\KKR} {R. Kallosh, J. Kumar and A. Rajaraman, hep-th/9712073}
\REF{\CKKTV} {P. Claus, R. Kallosh, J. Kumar, P. Townsend and A. Van Proeyen,
hep-th/9801206.}
\REF{\FF}  {S. Ferrara and C. Fronsdal,  hep-th/9712239.}
\REF{\2} {N. Itzhaki, J. M. Maldacena, J. Sonnenschein and S. Yankielowicz,
hep-th/9802042.}
\REF{\3} {M. Gunaydin and D. Minic,  hep-th/9802047.}
\REF{\4} {G. T. Horowitz and H. Ooguri, hep-th/9802116.}
\REF{\KS} {S. Kachru and E. Silverstein, hep-th/9802183.}
\REF{\6} {M. Berkooz,  hep-th/9802195.}
\REF{\65} {V. Balasubramanian and F. Larsen,  hep-th/9802198.}
\REF{\8} {J. M. Maldacena, hep-th/9803002.}
\REF{\9} {M. Flato and C. Fronsdal, hep-th/9803013.}
\REF{\LNV}  {A. Lawrence, N. Nekrasov and C. Vafa, hep-th/9803015.}
\REF{\newKleb} {S. S. Gubser, A. Hashimoto, I. R. Klebanov and M. Krasnitz,
hep-th/9803023.}
\REF{\CCD}{ L. Castellani, A. Ceresole, R. D'Auria, S. Ferrara, P. Fre'
and M. Trgiante, hep-th/9803039.}
\REF{\FFZ} {S. Ferrara, C. Fronsdal and A. Zaffaroni,  hep-th/9802203.}
\REF{\YAR}{O. Aharony, Y. Oz and Z. Yin, hep-th/9803051.}
\REF{\MIN}{S Minwalla, hep-th/9803053.}
\REF{\FZ} {S. Ferrara and A. Zaffaroni, hep-th/9803060.}
\REF{\DLP}{M. Duff, H. Lu and C. Pope, hep-th/9803061.}
\REF{\RO}{L. Romans, Phys. Lett. B154 (1985) 392.}
\REF{\GRW}{M. Gunaydin, L. Romans and N. Warner, Phys. Lett. B154 (1985) 268;
Nucl.
Phys. B272 (1985) 598.}
\REF{\LR}{R. Leigh and M. Rozali, hep-th/980368.}
\REF{\BKV}{M. Bershadsky, Z. Kakushadze and C. Vafa, hep-th/9803076.}
\REF{\EH}{E. Halyo, hep-th/9803077.}
\REF{\AR}{A. Rajaraman, hep-th/9803082.}
\REF{\HR}{G. Horowitz and S. Ross, hep-th/9803085.}
\REF{\FKPZ}{S. Ferrara, A. Kehagias, H. Partouche and A. Zaffaroni,
hep-th/9803109.}
\REF{\M}{J. Minahan, hep-th/9803111.}
\REF{\G}{J. Gomis, hep-th/9803119.}
\REF{\ED}{E. Witten, hep-th/9803131.}
\REF{\GU}{M. Gunaydin, hep-th/9803138.}
\REF{\OT}{Y. Oz and J. Terning, hep-th/9803167.}
\REF{\KRN} {H. J. Kim, L. J. Romans and P. van Nieuwenhuizen, Phys. Rev. D32
(1985)
389.}
\REF{\BCERS} {B. Biran, A. Casher, F. Englert, M. Rooman and P. Spindel,
Phys. Lett. {\bf 134B} (1984) 179.}
\REF{\GUN} {M. Gunaydin and N. P. Warner,  Nucl. Phys. B272 (1986) 99-124.}
\REF{\CENR} {A. Casher, F. Englert, H. Nicolai and M. Rooman,  Nucl. Phys. B243
(1984)
173-188.}
\REF{\NP}{B. E. W.  Nilsson and C. Pope, Class. Quant. Grav. 1 (1984) 499.}
\REF{\DUF}{M. Duff, H. Lu and C. Pope, hep-th/9704186.}

\singlespace
\rl{SU-ITP-98/16}
\rl{\today}
\rl{hep-th/9803193}
\pagenumber=0
\normalspace
\medskip
\bigskip
\titlestyle{\bf{ Supergravity on $AdS_{5/4} \times$ Hopf Fibrations and
Conformal Field Theories}}
\smallskip
\author{ Edi Halyo{\footnote*{e--mail address: halyo@dormouse.stanford.edu}}}
\smallskip
\centerline {Department of Physics}
\centerline{Stanford University}
\centerline {Stanford, CA 94305}
\smallskip
\vskip 2 cm
\titlestyle{\bf ABSTRACT}
We obtain three and four dimensional conformal field theories with less than
maximal supersymmetry by
using their supergravity duals. These supergravity theories are type II on
$AdS_5
\times CP^2$, IIA on $AdS_4 \times CP^3$, IIB on $AdS_5 \times S^5/Z_k$ and
$D=11$ supergravity on $AdS_4 \times S^7/Z_k$. They are obtained from
the spherically compactified ten and eleven dimensional theories by either
Hopf reduction or by winding the $U(1)$ fiber over the base.

\singlespace
\vskip 0.5cm
\endpage
\normalspace

\centerline{\bf 1. Introduction}

There has been great interest in the recently conjectured duality
between supergravity on $AdS_{d+1} \times S^n$ and conformal field theories
(CFT)
living on the boundary of the $AdS$ space which
is the Minkowski space in $d$ dimensions $M_d$[\MAL-\OT].
By this duality the isometry groups of the $AdS_{d+1}$ and $S^n$
which are $SO(d,2)$ and $SO(n+1)$ become the
conformal and R symmetries of the boundary theory. Since the spherical
compactifications of supergravity are maximally supersymmetric the
boundary superconformal field theory
SCFT has sixteen supercharges. Moreover, there is a correspondence
between the masses of bulk fields and the dimensions of the operators
in the boundary SCFT[\WIT]. For example, the tachyonic, massless and massive
scalars correspond
to relevant, marginal and irrelevant operators of the SCFT.
The $AdS_{d+1} \times S^n$ geometry is also the near horizon geometry of
nondilatonic branes of string and M theory, i.e. D3 branes of IIB,
M2 and M5 branes. Thus, one can see the SCFT on the boundary as the large
$N$ world--volume theories of these branes.
A nontrivial check of the conjecture is to match the dimensions of
marginal and relevant operators in the SCFT to the masses of Kaluza--Klein
towers
of fields in the bulk supergravity. This was shown for the
$AdS_5 \times S^5$ case in ref. [\WIT] and for $AdS_{4/7} \times S^{7/4}$ in
refs. [\YAR,\MIN,\LR,\EH]].
One can use this conjecture either to learn about SCFT by using the
spherical compactifications of supergravity or to learn about
nonperturbative aspects of string theory on $AdS$ spacetimes by using better
understood SCFT.

Spherical compactifications of supergravity are maximally supersymmetric
and therefore the boundary SCFT necessarily has sixteen supercharges. However,
the conjecture is believed to hold for any supersymmetry. SCFT with less than
sixteen supercharges have been investigated in refs.
[\KS,\LNV,\FZ,\FKPZ,\G,\OT]].
In all these cases
one reduces the supersymmetry by orbifolding the space transverse to the
boundary or brane. This leaves the $AdS$ part of the geometry intact resulting
in a CFT. The $S^n$ part of the geometry is orbifolded and depending on the
orbifold one gets different CFT with different amounts of supersymmetry.
In this paper, we obtain CFT with less than sixteen supercharges
by considering the transverse space $S^n$ to be a Hopf fibration. This can only
be done for the odd spheres
$S^5$ and $S^7$ which are $U(1)$ fibrations over $CP^2$ and
$CP^3$ respectively. We break supersymmetry either by reducing over
the $U(1)$ fiber, i.e. by Hopf reduction or by considering multiple windings
of the $U(1)$ over the base space, i.e. lens spaces.

The paper is organized as follows. In section 2, we review the aspects of
Hopf fibrations, Hopf reductions and lens spaces that we need in the
following. In section
3, we obtain three and four dimensional CFT with less than maximal
supersymmetry with their global symmetries and
marginal operators. This is done by Hopf reduction of
the IIB supergravity on $AdS_5 \times S^5$ and $D=11$ supergravity on
$AdS_4 \times S^7$. In sections 4 and 5 we do the same for three and four
dimensional CFT by considering the corresponding supergravities on lens
spaces. Section 6 is our conclusion.

\bigskip
\centerline{\bf 2. Hopf Reductions and Lens Spaces}

Any odd sphere $S^{2n+1}$ can be considered as a $U(1)$ fibration over $CP^n$.
The metric of the sphere becomes[\DLP]
$$d\Omega_{2n+1}=d\Sigma_{2n}^2+(dz+{\cal A})^2 \eqno(1)$$
where $d\Sigma_{2n}$ is the Fubini--Study metric of the $CP^n$ and the
Kaluza--Klein vector potential $\cal A$ has field strength $\cal F$ given by
${\cal F}=2J$ where $J$ is the Kahler form on $CP^n$. The coordinate on the
fiber is $z$ and it has period $4\pi$.

Now, both $AdS_5 \times S^5$ and $AdS_4 \times S^7$ can be written as Hopf
fibrations by subsituting eq. (1) into the metric instead of the spherical
parts $S^5$ and
$S^7$. For the $AdS_5 \times S^5$ case the metric becomes
(by taking $S^5$ to be a Hopf fibration over $CP^2$)
$$ds_{10}^2=ds^2(AdS_5)+{1 \over m^2} d\Sigma_4^2+{1 \over m^2}(dz+{\cal A})^2
\eqno(2)$$
where $1/m^2$ is the radius of the sphere and the AdS space related
to the cosmological constant of the AdS space and $d\Sigma_4^2$ is the metric
on $CP^2$. Since $\Omega_5=(dz+{\cal A}) \wedge \Sigma_4$ one can reduce
the ten dimensional metric on the $U(1)$ fiber and get a nine dimensional
metric[\DLP]
$$ds_9^2=ds^2(AdS_5)+{1 \over m^2} d\Sigma_4^2 \eqno(3)$$
This metric is a solution of the equations of motion of
the $D=9$ type II supergravity obtained from the $D=10$ IIB theory by Hopf
reduction.
(The solution also involves the dilaton and the five form in addition to the
metric but they
are not relevant to our discussion.)
Thus, it describes type II supergravity on $AdS_5 \times CP^2$. We are
interested
in the massless fields in the bulk since supersymmetry and the gauge group are
determined by the number of massless gravitini and massless gauge bosons.
These in turn determine the supersymmetry and R symetry of the boundary SCFT.
In addition, massless scalars correspond to marginal operators
in the boundary SCFT. The massless spectrum in the bulk of the $D=9$ theory
will be a truncation of that of the $D=10$ theory. The isometry of $S^5$
(R symmetry of the SCFT)  $SU(4)$  is reduced to $SU(3) \times U(1)$ which is
the
isometry of $U(1) \times CP^2$. The massless spectrum in the $D=9$ theory
given by the subset of the massless spectrum of $D=10$ theory which are zero
modes
of this reduction, i.e. they are neutral under the Kaluza--Klein gauge field
$U(1)$. As a result, the Hopf reduction will change the supersymmetry, gauge
symmetry and massless matter content of the bulk theory or equivalently
the supersymmetry, R symmetry and marginal operators of the four
dimensional boundary SCFT.

The $AdS^4 \times S^7$ case is very similar[\DUF]. Now $S^7$ is described as a
$U(1)$ fibration over $CP^3$ and one can Hopf reduce the $D=11$ supergravity
metric
for the $AdS_4 \times S^7$ over the fiber. This is a solution of $D=10$ IIA
theory and describes IIA on $AdS_4 \times CP^3$. The isometry of $S^7$ which is
$SO(8)$ is reduced to the isometry of $U(1) \times CP^3$ which is
$U(1) \times SU(4)$. Again truncating to the $U(1)$ neutral sector gives
the supersymmetry, R symmetry and marginal operators of the three dimensional
boundary SCFT.

Another way of getting SCFT with less than sixteen supercharges is to consider
supergravity on $AdS$ times a lens space denoted by $S^{2n+1}/Z_k$. These are
obtained by identifying the fiber coordinate of the $U(1)$ over $CP^{n}$ with
a period $1/k$ times that in the $S^{2n+1}$ case[\DLP]. These lens spaces are
equivalent to orbifolds of the space transverse to the boundary.
This can be seen by describing $S^{2n+1}$ by a unit sphere in $C^{n+1}$ with
complex coordinates $z_i, \quad i=1, \ldots,n+1$. Then the orbifold $z_i \to
e^{i \alpha}z_i$ with $\alpha=2\pi/k$ is precisely the lens space
$S^{2n+1}/Z_k$.
Again only a subset of the original states on $S^{2n+1}$ will remain in the
massless spectrum of $S^{2n+1}/Z_k$. These will be states with $U(1)$
charge[\DLP]
$$q={1 \over 2} kn \eqno(4)$$
where $n$ is any integer and $k$ is given by the $Z_k$ winding. Since
all gravitini, gauge bosons and scalars of the theory on the sphere carry
fixed charges under $U(1)$ different lens spaces (i.e. different $k$)
will result in different supersymmetry, R symmetry and marginal operators
for the three and four dimensional SCFT.

\bigskip
\centerline{\bf 3. Hopf Reduced Supergravity and Conformal Field Theories}

In this section we find the massless spectrum of $D=9$ type II supergravity
on $AdS_5 \times CP^2$ and of $D=10$ IIA supergravity on $AdS_4 \times CP^3$.
As explained above, we get these by Hopf reducing $D=10$ IIB supergravity on
$AdS_5 \times S^5$ [\RO,\GRW] and $D=11$ supergravity on $AdS_4 \times S^7$.
Using the conjectured duality between supergravity on $AdS$ spacetime and
SCFT on the boundary
these will give us four and three dimensional SCFT with less than maximal
supersymmetry.

Consider first $D=10$ IIB supergravity on $AdS_5 \times S^5$[\KRN]. The
isometry
group of $AdS_5 \times S^5$ is $SO(4,2) \times SU(4)$ where the first factor is
the conformal symmetry of the four dimensional boundary theory and the second
factor is the R symmetry. From the point of view of the bulk $SU(4)$ is the
gauge symmetry of the gauged supergravity with maximal (${\cal N}=8$)
supersymmetry.
Therefore, there are eight gravitini in the $4+ \bar 4$
and fifteen gauge bosons in the $15$ representations of $SU(4)$. In addition,
there are massless scalars in the bulk in the representations $1$, $45$ and
$105$ which correspond to marginal operators. In order to find out which
of these remain in the massles spectrum after Hopf reduction we
need to know the way
these representation decompose under the gauge symmetry of the $D=9$ theory
$SU(3) \times U(1)$
$$\eqalignno{&1 \to 1_0 &(5a) \cr
             &4 \to 3_{-1/2}+1_{3/2} &(5b) \cr
             &\bar 4 \to \bar 3_{1/2}+ 1_{-3/2} &(5c) \cr
             &15 \to 1_0+ 8_0+ 3_{-2}+ \bar 3_{2} &(5d) \cr
             &45 \to 3_4+ \bar 3_2+ 6_2+ 8_0+ 10_0+ 15_{-2} &(5e) \cr
             &105 \to  15^{\prime}_4+ \bar{15}^{\prime}_{-4}+24_{-2}+\bar
{24}_2+27_0                        &(5f)}$$
where the subscript denotes the $U(1)$ charges of the states. (The
normalization of the charges is chosen to aggree with supergravity liturature.)
Only the states
neutral under the $U(1)$ remain in the massless spectrum after Hopf reduction.
We see
that none of the gravitini remain so that there is no unbroken supersymmetry.
Of the gauge bosons only $1_0+ 8_0$ is left which means that the gauge group
is now $SU(3) \times U(1)$, precisely the group we expect from the isometry
of the $D=9$ theory. Among the scalars the remaining ones are in the
representations $1_0+ 8_0+ 10_0+27_0$.

Using the conjectured duality between supergravity on $AdS$ spaces and
CFT on the boundary we find that this decribes a $D=4$, ${\cal N}=0$ CFT
with a global $SU(3) \times U(1)$ symmetry and four marginal operators
in the $1_0+ 8_0+ 10_0+27_0$ representations.

Now we consider the second case which is $D=11$ supergravity on $AdS_4 \times
S^7$[\BCERS-\CENR]. The isometry group is $SO(3,2) \times SO(8)$ where the
first factor
is the conformal symmetry of the $2+1$ dimensional boundary theory and the
second one is the R symmetry. From the bulk point of view $SO(8)$ is the
gauge symmetry of the gauged supergravity with maximal supersymmetry.
Thus we have gravitini in one of the three representations $8_v,8_s,8_c$
(alternatively the gravitini are in the $8_s$ but there are three embeddings of
$SU(4) \times U(1)$ in $SO(8)$)
and gauge bosons in the $28$. The massless scalars and pseudoscalars are
in the $35_v$ and $35_s$ for $8_{v,c}$ gravitini. However for $8_s$ gravitini
the scalars and
pseudoscalars are both in the $35_v$.
The decomposition
of these under $SU(4) \times U(1)$ is
$$\eqalignno{&8_v \to 4_1+ \bar 4_1 &(6a) \cr
             &8_s \to 1_2+ 1_{-2}+ 6_0 &(6b) \cr
             &8_c \to 4_{-1}+ \bar 4_1 &(6c) \cr
             &28 \to 1_0+ 6_2+ 6_{-2}+ 15_0 &(6d) \cr
             &35_v \to 10_2+ \bar {10}_{-2}+ 15_0 &(6e) \cr
             &35_s \to 1_0+ 1_4+ 1_{-4}+ 6_2+ 6_{-2}+ 20^{\prime}_s &(6f) \cr
             &35_c \to 10_2+ \bar {10}_{-2}+15_0 &(6g)}$$
Again only the states neutral under the $U(1)$ remain after Hopf reduction.
There are three cases which correspond to the three possible representations
of the gravitini. The $v$ and $c$ cases are identical and we consider them
first. In this case, we find that there are no gravitini left so that there
is no supersymmetry in the bulk. The gauge group is broken down to $SU(4)
\times U(1)$ since the remaining gauge bosons are in $1_0+ 15_0$. The only
remaining
scalars (pseudoscalars) are in $15_0$ ($1_0$). This is the massless spectrum of
IIA supergravity on
$AdS_4 \times CP^3$. The corresponding boundary theory is a $D=3$, ${\cal N}=0$
CFT with global $SU(4) \times U(1)$ symmetry and two marginal operators in
the $1_0+15_0$.

Next we consider the other case with the gravitini in the $8_s$. We see that
in this case there are six gravitini in the $6_0$ which remain in the spectrum.
Therefore there is ${\cal N}=6$ supersymmetry in the bulk. The gauge group is
again $SU(4) \times U(1)$. The only massles scalars (and pseudoscalars)
are in the $15_0$. On the
boundary this gives a $D=3$, ${\cal N}=6$ SCFT with R symmetry $SU(4) \times
U(1)$ and two marginal operators in  the $15_0$.

\bigskip
\centerline{\bf 4. IIB Supergravity on $AdS_5 \times$ Lens Space}

In this section, we consider IIB supergravity on $AdS_5 \times S^5/Z_k$ [\DLP]
where
the second factor denotes a lens space described in section 2. The winding
of the $U(1)$ fiber over $CP^2$ is another way of breaking supersymmetry and
the gauge group of the bulk theory. This happens because only a subset of the
states of IIB supergravity on $AdS_5 \times S^5$ survive the charge condition
given in eq. (4). As mentioned previously, different winding numbers $k$ will
result in different superymmetries and gauge groups. For concreteness
we consider the cases with $k=2,3$ explicitly and make some general
observations
about the cases with larger $k$.

$k=2$: In this case, the charge condition for the surviving states is
$q=n$ where $n$ is any integer. We find that all gravitini are projected out
so that there is no supersymmetry. The gauge group in the bulk becomes
$SU(3) \times SU(2)^2 \times U(1)$ since all gauge bosons remain in the
spectrum
but in the representations $1_0+ 3_{-2}+ 3_2+ 8_0$. All the massless scalars
also remain but are now decomposed into $1_0$ the singlet,
$3_4+ \bar 3_2+ 6_2+ 8_0+ 10_0+ 15_{-2}$ coming from the $45$
and $15^{\prime}_4+ \bar{15}^{\prime}_{-4}+24_{-2}+\bar {24}_2+27_0 $ coming
from the
$105$.
By the duality this corresponds to a $D=4$, ${\cal N}=0$ CFT with a global
$SU(3) \times SU(2)^2 \times U(1)$ group and thirteen marginal operators in the
representations given above.

$k=3$: Now only states with $q=3n/2$ survive. As a result, there are two
remaining gravitini giving ${\cal N}=2$ supersymmetry in the bulk. The gauge
group is $SU(3) \times U(1)$ and the remaining scalars are in the $1_0$,
$8_0+10_0$ and $27_0$ representations. On the boundary this gives a $D=4$,
${\cal N}=1$
SCFT with four marginal operators. This is exactly the same SCFT obtained
in ref. [\KS,\OT] by orbifolding the the space transverse to the D3 branes.

By now it is easy to see some general features of the boundary CFT depending on
the
winding number $k$.
We see that for $k>3$ all gravitini are projected out so that these lead to
four dimensional
CFT without supersymmetry. For $k=4$ the gauge group is $SU(3) \times
SU(2)^2 \times U(1)$ whereas for $k>4$ it is $SU(3) \times U(1)$. Four
marginal operators always survive since there are three massless scalars
with vanishing $U(1)$ charge: the singlet $1_0$, $8_0+10_0$ coming from the
$45$ and $27_0$ coming from the $105$.

\bigskip
\centerline{\bf 5. $D=11$ Supergravity on $AdS \times$ Lens Space}

In this section, we consider $D=11$ supergravity on $AdS_4 \times$ lens space
which we denote $S^7/Z_k$[\NP,\DUF]. Now $S^7$ is taken to be a $U(1)$
fibration over
$CP^3$ and $k$ again denotes the winding of the fiber over the base.
Again for different $k$ we will find that different subsets of the massless
sector of $D=11$ supergravity on $AdS_4 \times S^7$ survive resulting in
different CFT on the boundary. Below we consider the cases for $k=2,3$
explicitly and make some general observations about other $k>3$.

$k=2$: For $k=2$ the cases with gravitini in $8_v$ (or $8_c$) and $8_s$ lead to
the same results. The charge
condition is $q=n$ and we find that all gravitini remain in the spectrum and
there is maximal supersymmetry in the bulk. We also find that all the gauge
bosons remain in the spectrum; however the gauge group is broken down to $SU(4)
\times SO(4)^2
\times U(1)$. In addition, all massless scalars (and pseudoscalars) remain but
now in the
$10_2+ \bar {10}_{-2}+15_0$ (and $1_4+1_{-4}+1_0+ 6_2+ 6_{-2}+ 20^{\prime}_s$)
representation for the gravitini in $8_{v,c}$ case and
both in the $10_2 +\bar {10}_{-2}+15_0$ for the gravitini in the
$8_s$ case. We find that none of the massless states
are projected out for $k=2$ since all $U(1)$ charges are integer in eq. (6).
This leads on the boundary to $D=3$, ${\cal N}=8$ SCFT with the $SU(3) \times
SO(4)^2 \times U(1)$ R symmetry.

$k=3$: For $k=3$ the charge condition is $q=3n/2$ and there will be a
difference
between the two cases in which the gravitini belong to different
representations. In the case of $8_{v,c}$ gravitini, all gravitini are
projected out so that the bulk theory is not supersymmetric. The gauge group is
$SU(4) \times
U(1)$ in the bulk. The only remaining massless scalar (pseudoscalar) is in the
$15_0$ ($1_0$)
representation. Thus we get on the boundary a $D=3$ ${\cal N}=0$ CFT with
global $SU(4) \times U(1)$ symmetry and two marginal operators.
In the case of $8_s$ gravitini, we find that six gravitini in the $6_0$
remain so that there is ${\cal N}=6$ supersymmetry in the bulk. The gauge
group is exactly as in the other case but now both scalars and pseudoscalars
are in the $15_0$.
This leads to a $D=3$, ${\cal N}=6$ SCFT with $SU(4) \times U(1)$ R symmetry
with two
marginal operators.

For larger $k$ some general features emerge from the $U(1)$ charges given
in eq. (6).
We see that in the case of $8_{v,c}$ gravitini for $k>3$ there is no
supersymmetry whereas in the case of $8_s$ gravitini for $k=4$ there is
maximal supersymmetry and for $k>4$ there is ${\cal N}=6$ supersymmetry
in the CFT. In both cases the global symmetry group is $SU(4) \times SO(4)^2
\times U(1)$ for $k=4$ and $SU(4) \times U(1)$ for $k>4$. We find that for all
$k$ there
are at least two marginal operators either in $1_0$ for $8_{v,c}$ gravitini or
in $15_0$
for $8_s$ gravitini. We find that for all $k$ there are at least two marginal
operators since
$35_{v,c}$ includes $15_0$ and $35_s$ includes $1_0$ which cannot be projected
out.

\medskip
\centerline{\bf 6. Conclusions}

In this paper, we obtained four and three dimensional CFT with less than
maximal supersymmetry using the conjectured duality between supergravity
on $AdS$ spaces and CFT on the boundary. We found the CFT dual to
$D=9$ type II supergravity on $AdS_5 \times CP^2$ and $D=10$ IIA supergravity
on $AdS_4 \times CP^3$. These were obtained by Hopf reduction of IIB on
$AdS_5 \times S^5$ and $D=11$ supergravity on $AdS_4 \times S^7$. The
odd spheres $S^{5,7}$ were considered to be $U(1)$ fibrations over $CP^{2,3}$
and the theories were dimensionally reduced over the fiber. Clearly this
method is only useful for spacetimes which contain an odd sphere. For example,
this method cannot be applied to the third well--known case of $D=11$
supergravity on $AdS_7 \times S^4$. We also considered CFT which are dual to
$D=11$ and IIB
supergravity compactified on lens spaces where the $U(1)$ fiber is
wound over the base $CP^n$ $k$ times. We saw that the number of
supersymmetries, the
global symmetry and the number of marginal opeartors in the boundary CFT
depend on the winding number $k$.

We obtained a number of CFT with less than maximal or no supersymmetry
in both three and four dimensions. In particular we were able to find
nonsupersymmetric
CFT fairly easily either by Hopf reduction or by winding the $U(1)$ fiber. This
is due to the fact
that all gravitini have fixed and nonzero $U(1)$ charges and therefore in Hopf
reduction
and on most lens spaces they are projected out completely. We also found that
all these  CFT  have marginal operators. The massless scalars
decompose into states which include $U(1)$ neutral states which cannot be
projected out.
In this paper we were only interested in the marginal operators of the CFT
which are related
to the massles bulk fields in supergravity. However, one can easily do the same
analysis
for the tachyonic fields which will give the relevant operators in the CFT
discussed above.

Even though we were able to find CFT with less than maximal supersymmetry
by using the duality to supergravity on $AdS$ spacetimes we could not check our
results by direct
knowledge of the CFT on the boundary. In particular we do not know a
construction of these CFT by using D branes or M branes (except the $S^5/Z_k$
case which is equivalent to orbifolded D3 brane theories).
However, since the original supergravities are dual to world--volume theories
of D3 or M2 branes before Hopf reduction or winding the fiber one expects
that such a description is possible. It would be interesting to find these
descriptions and explicitly check the conjectured duality in the cases
considered in this paper.

\medskip
\centerline{\bf Acknowledgements}
We would like to thank Arvind Rajaraman and Eva Silverstein for very useful
discussions.

\vfill

\refout
\bye

\end